\documentclass[a4paper,11pt]{article}
\usepackage[dvips]{graphicx}
\usepackage{amssymb}
\usepackage{amsmath}

\usepackage{cite}

\def\G{{\bf G}}

\input arrow

\begin{document}

\title{Graded Geometric Structures Underlying F-Theory Related Defect Theories}
\author{V. K. Oikonomou\thanks{
voiko@physics.auth.gr}\\
Max Planck Institute for Mathematics in the Sciences\\
Inselstrasse 22, 04103 Leipzig, Germany} \maketitle

\begin{abstract}
In the context of F-theory, we study the related eight dimensional
super-Yang-Mills theory and reveal the underlying supersymmetric
quantum mechanics algebra that the fermionic fields localized on
the corresponding defect theory are related to. Particularly, the
localized fermionic fields constitute a graded vector space, and
in turn this graded space enriches the geometric structures that
can be built on the initial eight-dimensional space. We construct
the implied composite fibre bundles, which include the graded
affine vector space and demonstrate that the composite sections of
this fibre bundle are in one-to-one correspondence to the sections
of the square root of the canonical bundle corresponding to the
submanifold on which the zero modes are localized.
\end{abstract}

\section*{Introduction}

String theory has proven to be the most promising theory towards
the unified description of all forces and matter in nature.
Particularly, it encompasses in its theoretical framework gravity
and a large number of field theoretic features such as,
supersymmetry, chiral matter and spontaneous symmetry breaking.
The appealing attribute of string theory is that it can provide
consistent UV completion of many field theoretic models because it
can accommodate quite successfully supersymmetric grand unified
theories. M-theory embodies all the different string theories that
describe independently various features of the UV completions of
the Standard Model (SM hereafter), with the various branches of
M-theory being connected with dualities, a strong tool towards a
non-perturbative description. However, certain branches of
M-theory prove to be more efficient in realizing SM
phenomenological features, than others. Particularly, type IIB
string theory and the strongly coupled version of it, F-theory
(for an important stream of papers on F-theory see
\cite{vafa1,vafa2,vafa3,freview,freview1,freview2,fzero,fzero1,fzero2,fgut,fgut1,fgut2,fgut3,fgut4,fvarious,fvarious1,fvarious2,fvarious3,fvarious4,fvarious5,fvarious6,fvarious7,fvarious8,fvarious9,fvarious10}
and for reviews on F-theory see \cite{freview,freview1,freview2})
embody many phenomenological features of the SM. The most
appealing feature of these theories is that these allow gauge
interactions, existence and propagation of matter in a way that is
independent from the full vacuum configuration, with the last
being one of the most difficult problems in string theory, because
there is a huge number of possible vacua. In the type
IIB--F-theory framework, gauge interactions, such as Yukawa
couplings, and chiral matter are localized on sub-manifolds of the
total space, and thus depend only on the local geometry to some
extend. $D$7 branes are necessary to describe chiral matter and in
type IIB theories this type of matter is realized on the
intersection of such objects. The use of these low dimensional
D-branes, enables one to describe low energy physics in a
bottom-up way, because only local configurations of these $D$7
branes are considered and these are localized, as we already
mentioned, to some region of the compact dimensions. Thus, the
IIB--F-theory description is a successful to some extend,
bottom-up approach to the problem of finding an appropriate low
energy string compactification. In the F-theory context, gravity
can be decoupled from SM physics, since $D$7 branes that contain a
GUT group, wrap certain classes of complex surfaces $S$ with four
real dimensions, and if the infinite volume limit  of these
surfaces is taken, the SM physics is obtained but decoupled from
gravity. Particularly, an $N=1$ supersymmetric GUT theory with a
singularity enhanced gauge group can either result from M-theory
on $G_2$-holonomic seven dimensional manifolds, or from F-theory
compactified on elliptic Calabi-Yau fourfolds. In the case of
F-theory compactifications, the K\"{a}hler geometry provides
freedom to use numerous geometric techniques in order to
accommodate many phenomenological features of the Standard Model
into the theoretical outcomes of such theories. Moreover, non-zero
Yukawa couplings exist for fields that reside on intersecting $D$7
branes. Actually, most of the most interesting SM phenomenology
can be deduced from an eight dimensional super-Yang-Mills-Higgs
theory living on $D$7 branes. The Yukawas are obtained from the
overlap of the chiral zero modes on the seven-branes (see
\cite{fgut,fgut1,fgut2,fgut3,fgut4} and
\cite{fzero,fzero1,fzero2}).

\noindent F-theory on a local Calabi-Yau fourfold is described at
low energies by the twisted super-Yang-Mills theory living on the
worldvolume of the seven brane that wraps $S\times R^{3,1}$. Let a
seven brane wrapping the complex dimension two surface $S$,
intersect with another seven brane that wraps $S'$. We denote
their intersection $\Sigma=S\cap S'$, a complex dimension one
curve. The physics of the charged fields that reside on the
intersection $\Sigma \times R^{3,1}$ can be consistently described
by a twisted six dimensional defect theory coupled to the bulk
theories $S\times R^{3,1}$ and $S'\times R^{3,1}$. We assume that
$\Sigma$ is irreducible, that is, it does not consist of several
components. As we already mentioned, light degrees of freedom are
localized on the manifold $\Sigma \times R^{3,1}$. It is exactly
these degrees of freedom that are described by an effective defect
theory coupled to the bulk  $S\times R^{3,1}$ and $S'\times
R^{3,1}$ super-Yang Mills theories. The structure of the defect
theory $\Sigma \times R^{3,1}$ can be determined directly from the
$S\times R^{3,1}$ super-Yang Mills theory itself. Starting from a
seven brane with world volume gauge group $\Gamma_S$, wrapping the
complex surface $S$, in order to obtain an intersection of seven
branes, we can allow the scalar field $\phi$ belonging to the
$N=1$ supersymmetric $S\times R^{3,1}$ theory to have a
non-trivial holomorphic vacuum expectation value. The scalar field
vanishes on some points of the curve $\Sigma$. On the points in
which $\phi$ vanishes, the bulk gauge group $G_S$ remains
unbroken, but away from the singularity, the gauge group becomes
some $\Gamma_S \subset G_S$, times a $U(1)$. The structure of the
defect theory can be captured by examining what happens in the
$N=1$ supersymmetric $S\times R^{3,1}$ theory, when $\phi$
acquires a non-trivial holomorphic expectation value. The defect
theory on $\Sigma$ can be viewed from the point of view of the
complex-dimension two surface $S$, as a global string associated
to the vanishing of the holomorphic mass term that is induced by
the scalar $\phi$. Hence, we expect (and this is exactly the case)
to find bosonic and fermionic zero modes trapped along $\Sigma$,
thus resulting to massless, chiral matter on $R^{3,1}$. Thereby,
identifying which fermionic zero modes are trapped in the
submanifold $\Sigma$, simultaneously determines which bosonic zero
modes are localized.

\noindent In this paper the focus is on exactly these fermionic
zero modes that are trapped on the complex dimension one curve
$\Sigma $, with $\Sigma$ being considered as a defect of the
theory on $S$. We establish the result that these fermionic zero
modes are associated to an one dimensional $N=2$ supersymmetric
quantum mechanics (SUSY QM hereafter) algebra
\cite{susyqm,susyqm1,susyqm2,susyqm3,susyqm4,susyqm5,susyqm6,susyqm7,susyqm8,susyqm9,susyqm10,susyqm11,susyqm12,susyqm13}.
Particularly, the zero modes localized on $\Sigma $ are in
bijective correspondence with the vectors of the graded Hilbert
space that describes the SUSY QM vector space of quantum states
(for an similar case related to superconducting strings and to
Chern-Simons gauge theories in $(2+1)$-dimensions, see
\cite{oikonomou,oikonomou1}). Moreover, the underlying unbroken
SUSY QM algebra makes the total eight dimensional space $S\times
R^{3,1}$ a graded manifold $(X,\mathcal{A})$, with body $X\equiv
S\times R^{3,1}$ and structure sheaf $\mathcal{A}$. Therefore the
total space $S\times R^{3,1}$ has a potentially rich variety of
geometric structures that can be constructed on it. We are
particularly interested on the composite fibre bundles that can be
constructed on this space, in which the graded manifold belongs.
As we shall see, from all the sections of the $G_S$-twisted spin
bundle upon $S\times R^{3,1}$, only those trapped on $\Sigma$ are
connected to the SUSY QM algebra. This implies that the covariant
differential of the $\G_S$-twisted spin bundle over $S\times
R^{3,1}$ is reducible to some covariant differential of a
composite fibre bundle over some subbundle of $S\times R^{3,1}$,
with this bundle containing the graded bundle in its substructure.
The bundle reducibility is realized in terms of the corresponding
connections of the composite fibre bundles. In the following
sections we shall study these issues in detail, and reveal the
SUSY QM generated underlying geometric structures.

\noindent This paper is organized as follows. In section 1 we
present the details of the $S\times R^{3,1}$ super-Yang Mills
theory, and find which modes of the initial eight-dimensional
theory are trapped to move along the submanifold $\Sigma$. In
addition, we substantiate the result that these modes are
associated to an unbroken SUSY QM algebra, which we describe
briefly. We also study how holomorphic perturbations to the
Euclidean metric affect the SUSY QM algebra. In section 2 we
present the additional geometric structures over the space
$S\times R^{3,1}$, which is due to the  unbroken $N=2$ SUSY QM
algebra and discuss the features of the fermionic geometric
structures over $S\times R^{3,1}$. The concluding remarks follow
in the end of the paper.

\section{Defect Theory on the Submanifold $\Sigma$, Localized Zero Modes and $N=2$ SUSY QM}

In this section we shall present the details of the $S\times
R^{3,1}$ super-Yang Mills theory, working locally in $\Sigma
\subset S$. We adopt the notation and conventions of
\cite{vafa1,vafa2,vafa3}. Parameterizing the space $S$ with two
holomorphic coordinates $(s^1,s^2)$, and since the field $\phi$
vanishes on $\Sigma$, the field $\phi$ and its conjugate,
$\bar{\phi}$, can take the form:
\begin{equation}\label{phifield}
\phi=t s^2\mathrm{d}s^1\wedge \mathrm{d}s^2,{\,}{\,}{\,}\bar{\phi}=t \bar{s}^2\mathrm{d}\bar{s}^1\wedge \mathrm{d}\bar{s}^2
\end{equation}
Consequently, the complex dimension one curve $\Sigma$, is defined
to be the submanifold of $S$, for which $s^2 =0$. Therefore, $s^1$
parameterizes tangent directions on the curve, while $s^2$ normal
directions to $\Sigma$. It is important for the geometric
constructions we shall present in the next section, to
en-visualize this geometric structure, in the way that $s^2$ shows
directions that belong to $S$ but not in $\Sigma$. The fact that
the canonical bundle of the curve $\Sigma$ consists of the
differential $\mathrm{d}s^1$, has important consequences, as we
shall see. The part of the whole eight-dimensional super-Yang
Mills action, that is relevant for the fermionic equations of
motion is:
\begin{align}\label{bilinear}
& I_F=\int_{S\times R^{3,1}}\mathrm{d}x^4\mathrm{Tr}\Big{(}\chi^a
\wedge\bar{\partial_A}\psi_a+\bar{\chi_a}
\wedge\partial_A\bar{\psi^a}+2{\,}i\sqrt{2}{\,}\omega \wedge\partial_A\eta^a
\wedge \psi_a \\ \notag & +2{\,}i\sqrt{2}{\,}\omega \wedge\bar{\partial_A}\bar{\eta_a}
\wedge \bar{\psi^a}-\frac{1}{2}\bar{\psi_a}\wedge [\bar{\phi},\bar{\psi^a}]+\frac{1}{2}\psi^a\wedge [\phi,\psi_a]+\sqrt{2}{\,}\bar{\eta_a}
[\bar{\phi},\chi_a]+\sqrt{2}{\,}\eta^a
[\phi,\bar{\chi^a}]\Big{)}
\end{align}
Thereby, the fermionic equations of motion read,
\begin{align}\label{eqmotion1}
& \bar{\partial_A}\psi^a-\sqrt{2}[\bar{\phi},\eta^a]=0,{\,}{\,}{\,}{\,}{\,}{\,}{\,}{\,}{\,}{\,}{\,}{\,}{\,}{\,}{\,}\partial_A\bar{\psi^a}+\sqrt{2}[\phi,\bar{\eta^a}]=0 \\ \notag &
\omega\wedge \bar{\partial_A}\eta^a+\frac{i\sqrt{2}}{4}[\bar{\phi},\bar{\psi^a}]=0,{\,}{\,}{\,}\omega\wedge \partial_A\bar{\eta^a}++\frac{i\sqrt{2}}{4}[\phi,\psi^a]=0=0
\end{align}
for the $(\eta,\psi)$ fermion pairs, while for the $(\chi,\psi)$ pair, we have:
\begin{align}\label{eqmotion12}
& \omega\wedge \partial_A\psi^a+\frac{i}{2}[\bar{\phi},\chi^a]=0,{\,}{\,}{\,}{\,}{\,}{\,}{\,}{\,}{\,}{\,}{\,}{\,}{\,}{\,}{\,}\omega\wedge \bar{\partial_A}\bar{\psi^a}-\frac{i}{2}[\phi,\bar{\chi^a}]=0 \\ \notag &
\bar{\partial_A}\chi^a-[\phi,\psi^a]=0,{\,}{\,}{\,}{\,}{\,}{\,}{\,}{\,}{\,}{\,}{\,}{\,}{\,}{\,}{\,}{\,}{\,}{\,}{\,}{\,}{\,}{\,}{\,}{\,}{\,}{\,}{\,}\partial_A\bar{\chi^a}-[\bar{\phi},\bar{\psi^a}]=0=0
\end{align}
The partial derivatives are $\partial_A=\frac{\partial}{\partial s^2}$, and $\omega$ is the canonical Euclidean form near $\Sigma$:
\begin{equation}\label{dfredd}
\omega =\frac{i}{2}\large{(}\mathrm{d}s^1\wedge
\mathrm{d}\bar{s}^1+\mathrm{d}s^2\wedge \mathrm{d}\bar{s}^2\large{)}
\end{equation}
With $\omega$ being of this form, no gravitational back-reaction is taken to account. Later on we shall perturb this form in order to study metric back-reaction effects on the SUSY QM algebra. From all the equations appearing in (\ref{eqmotion1}) and (\ref{eqmotion12}), only those of relation (\ref{eqmotion12}) give normalized localized solutions near $\Sigma$, with the solutions exponentially decaying as:
\begin{equation}\label{expdec}
\Psi \sim \exp (-|s^2|^2)
\end{equation}
Substituting $\phi$ from equation (\ref{phifield}), the equations of motion (\ref{eqmotion12}) take the general form (ignoring irrelevant constants):
\begin{align}\label{basiceqm}
&\frac{\partial \Psi}{\partial s^2}+\bar{s}^2\tilde{\Psi}=0 \\ \notag &
\frac{\partial \tilde{\Psi}}{\partial \bar{s}^2}+s^2\Psi=0
\end{align}
with $\Psi$ and $\tilde{\Psi}$ denote any pair $\psi$ and $\chi$
respectively, or their conjugates. Particularly, as discussed in
\cite{vafa1,vafa2,vafa3}, each of the fermions $\chi$ and $\psi$
and their conjugates, have exactly one zero mode that behaves as
(\ref{expdec}). Since all our results hold for both pairs of
fermions, we focus for simplicity on the fermions $(\psi,\chi)$.

\noindent Thereby, we can form the operator $\mathcal{D}_{F}$,
corresponding to the equations of (\ref{basiceqm}) (with $\partial_2=\partial/\partial s^2$),
\begin{equation}\label{susyqmrn567m}
\mathcal{D}_{F}=\left(%
\begin{array}{cc}
 \partial_2 & \bar{s}^2
 \\ s^2 & \bar{\partial}_2\\
\end{array}%
\right)
\end{equation}
acting on the vector:
\begin{equation}\label{ait34e1}
|\Psi_{F}\rangle =\left(%
\begin{array}{c}
  \Psi \\
  \tilde{\Psi} \\
\end{array}%
\right).
\end{equation}
Therefore, the equations (\ref{basiceqm}) can be written as follows:
\begin{equation}\label{transf}
\mathcal{D}_{F}|\Psi_{F}\rangle=0
\end{equation}
The solutions of equation (\ref{transf}) correspond to the zero modes of the
operator $\mathcal{D}_{F}$. Since we know that there is exactly one zero mode for each $\chi$ and $\psi$, we have that
\begin{equation}\label{dimeker}
\mathrm{dim}{\,}\mathrm{ker}\mathcal{D}_{F}=1
\end{equation}
Moreover, the adjoint of the operator $\mathcal{D}_{F}$,
namely $\mathcal{D}_{F}^{\dag}$, is:
\begin{equation}\label{eqndag}
\mathcal{D}_{F}^{\dag}=\left(%
\begin{array}{cc}
 \bar{\partial}_2 & \bar{s}^2
 \\ s^2 & \partial_2\\
\end{array}%
\right)
\end{equation}
The corresponding kernel of the adjoint operator is null, that is:
\begin{equation}\label{dimeke1r11}
\mathrm{dim}{\,}\mathrm{ker}\mathcal{D}_{F}^{\dag}=0
\end{equation}
Since the kernel of the operator $\mathcal{D}_{F}$ is finite, this
property renders it a Fredholm operator. Using the operator
$\mathcal{D}_{F}$, we can construct an unbroken $N=2$, $d=1$
supersymmetry for the system $(\chi,\psi)$ of fermions. Indeed, we
can form the supercharges and the quantum Hamiltonian of the
$N=2$, $d=1$ SUSY algebra:
\begin{equation}\label{s7}
\mathcal{Q}_{F}=\bigg{(}\begin{array}{ccc}
  0 & \mathcal{D}_{F} \\
  0 & 0  \\
\end{array}\bigg{)},{\,}{\,}{\,}\mathcal{Q}^{\dag}_{F}=\bigg{(}\begin{array}{ccc}
  0 & 0 \\
  \mathcal{D}_{F}^{\dag} & 0  \\
\end{array}\bigg{)},{\,}{\,}{\,}\mathcal{H}_{F}=\bigg{(}\begin{array}{ccc}
 \mathcal{D}_{F}\mathcal{D}_{F}^{\dag} & 0 \\
  0 & \mathcal{D}_{F}^{\dag}\mathcal{D}_{F}  \\
\end{array}\bigg{)}
\end{equation}
These operators satisfy the $N=2$, $d=1$ SUSY QM
algebra:
\begin{equation}\label{relationsforsusy}
\{\mathcal{Q}_{F},\mathcal{Q}^{\dag}_{F}\}=\mathcal{H}_{F}{\,}{\,},\mathcal{Q}_{F}^2=0,{\,}{\,}{\mathcal{Q}_{F}^{\dag}}^2=0
\end{equation}
There is an operator $W$, the Witten parity, that commutes with
the total Hamiltonian and anti-commutes with the supercharges,
\begin{equation}\label{s45}
[\mathcal{W},\mathcal{H}_{F}]=0,{\,}{\,}\{\mathcal{W},\mathcal{Q}_{F}\}=\{\mathcal{W},\mathcal{Q}_{F}^{\dag}\}=0
\end{equation}
Moreover, $\mathcal{W}$, satisfies the following
identity,
\begin{equation}\label{s6}
\mathcal{W}^{2}=1
\end{equation}
The total Hilbert space of the supersymmetric quantum mechanical system,
$\mathcal{H}$, is rendered an $Z_2$ graded vector space, with the
grading provided by the operator $\mathcal{W}$, which is actually an involution
operator and decomposes as:
\begin{equation}\label{fgjhil}
\mathcal{H}=\mathcal{H}^+\oplus \mathcal{H}^-
\end{equation}
with the vectors belonging to the subspaces
$\mathcal{H}^{\pm}$, classified to even and odd parity states according to their Witten parity, that is:
\begin{equation}\label{shoes}
\mathcal{H}^{\pm}=\mathcal{P}^{\pm}\mathcal{H}=\{|\psi\rangle :
\mathcal{W}|\psi\rangle=\pm |\psi\rangle \}
\end{equation}
The corresponding Hamiltonians of the $Z_2$ graded
spaces are:
\begin{equation}\label{h1}
{\mathcal{H}}_{+}=\mathcal{D}_{F}{\,}\mathcal{D}_{F}^{\dag},{\,}{\,}{\,}{\,}{\,}{\,}{\,}{\mathcal{H}}_{-}=\mathcal{D}_{F}^{\dag}{\,}\mathcal{D}_{F}
\end{equation}
In our case, the operator $\mathcal{W}$, has the corresponding representation:
\begin{equation}\label{wittndrf}
\mathcal{W}=\bigg{(}\begin{array}{ccc}
  1 & 0 \\
  0 & -1  \\
\end{array}\bigg{)}
\end{equation}
The eigenstates of the operator $\mathcal{P}^{\pm}$, that is,
$|\psi^{\pm}\rangle$ (positive and negative parity eigenstates),
satisfy the following relation:
\begin{equation}\label{fd1}
P^{\pm}|\psi^{\pm}\rangle =\pm |\psi^{\pm}\rangle
\end{equation}
Using (\ref{wittndrf}) for the Witten parity operator,
the parity eigenstates are the vectors,
\begin{equation}\label{phi5}
|\psi^{+}\rangle =\left(%
\begin{array}{c}
  |\phi^{+}\rangle \\
  0 \\
\end{array}%
\right),{\,}{\,}{\,}
|\psi^{-}\rangle =\left(%
\begin{array}{c}
  0 \\
  |\phi^{-}\rangle \\
\end{array}%
\right)
\end{equation}
with $|\phi^{\pm}\rangle$ $\in$ $\mathcal{H}^{\pm}$. It is easy to verify that:
\begin{equation}\label{fdgdfgh}
|\Psi_{F}\rangle =|\phi^{-}\rangle=\left(%
\begin{array}{c}
  \Psi \\
  \tilde{\Psi} \\
\end{array}%
\right)
\end{equation}
Therefore, the corresponding even and odd parity SUSY QM states
are:
\begin{equation}\label{phi5}
|\psi^{+}\rangle =\left(%
\begin{array}{c}
  |\Psi_{F}\rangle \\
  0 \\
\end{array}%
\right),{\,}{\,}{\,}
|\psi^{-}\rangle =\left(%
\begin{array}{c}
  0 \\
  |\Psi_{F}\rangle \\
\end{array}%
\right)
\end{equation}
Supersymmetry is considered unbroken if the Witten index is a non-zero integer. The Witten
index when the operators are Fredholm operators is:
\begin{equation}\label{phil}
\Delta =n_{-}-n_{+}
\end{equation}
with $n_{\pm}$ the number of zero modes of ${\mathcal{H}}_{\pm}$
in the subspace $\mathcal{H}^{\pm}$.

\noindent When $\Delta$ is zero and also if
$n_{+}=n_{-}=0$, then supersymmetry is broken. Nevertheless, if $n_{+}=
n_{-}\neq 0$ the system retains unbroken supersymmetry.

\noindent The Fredholm index of the operator $\mathcal{D}_{F}$ is connected to the Witten index, as:
\begin{align}\label{ker1}
&\Delta=\mathrm{dim}{\,}\mathrm{ker}
{\mathcal{H}}_{-}-\mathrm{dim}{\,}\mathrm{ker} {\mathcal{H}}_{+}=
\mathrm{dim}{\,}\mathrm{ker}\mathcal{D}_{F}^{\dag}\mathcal{D}_{F}-\mathrm{dim}{\,}\mathrm{ker}\mathcal{D}_{F}\mathcal{D}_{F}^{\dag}=
\\ \notag & \mathrm{ind} \mathcal{D}_{F} = \mathrm{dim}{\,}\mathrm{ker}
\mathcal{D}_{F}-\mathrm{dim}{\,}\mathrm{ker}
\mathcal{D}_{F}^{\dag}
\end{align}
In our case, the Witten index is:
\begin{equation}\label{fnwitten}
\Delta =1
\end{equation}
Hence, some of the fermions of the eight dimensional super-Yang Mills theory at hand, are related to the graded vectors of an underlying $N=2$, $d=1$ unbroken supersymmetry. We could say that this
is in virtue of the fact that the initial system has an $N=1$ spacetime supersymmetry, so the Hilbert space of
the zero modes states inherits an $N=2$, $d=1$
supersymmetric quantum algebra. But this is not true since
global spacetime supersymmetry in $d>1$ dimensions and $d=1$ supersymmetry (supersymmetric quantum mechanics), are
not the same both conceptually and quantitatively. Indeed, the SUSY QM supercharges do not generate any
transformations between fermionic and bosonic fields. Moreover, these
supercharges provide a $Z_2$-grading to the corresponding Hilbert space. It is obvious that the $Z_2$-grading that the SUSY QM algebra provides to some of the fermion states, has nothing to do with the initial supersymmetry. But this grading has a direct impact on the geometry of the associated bundles corresponding to the space $S\times R^{3,1}$. Before we present the geometric implications of the underlying SUSY QM algebra, let us see whether this algebra remains intact under holomorphic gravitational back-reaction.

\noindent So far we supposed that the canonical form for the
metric that describes $S$ had the Euclidean form. It is
interesting to examine what is the impact of a deformation of the
metric on the SUSY QM algebra we presented earlier. The complex
dimension two surface $S$ is more like a base space of the
Calabi-Yau threefold and not a divisor \cite{fzero,fzero1,fzero2}.
Since the approach we adopted describes $S$ locally and not
globaly, there is no consistent way to know the full form of the
metric that fully describes the base space $S$. The freedom that
stems from this local approach, enables us to use a deformed form
of the Euclidean metric we used in order to describe $S$. As we
already mentioned, the canonical Euclidean form perfectly
describes a gravity decoupled Super Yang-Mills theory. We shall
exploit the freedom in selecting the local metric, and use a
metric that incorporates some gravitational back-reaction of the
complex surface $S$, which is of the form:
\begin{equation}\label{metric}
\mathrm{d}s^2=\big{(}1+\epsilon f_1(s^1)\big{)}
\mathrm{d}s^1\otimes\mathrm{d}\bar{s}^11+\big{(} 1+\epsilon
f_2(s^2)\big{)}\mathrm{d}s^2\otimes\mathrm{d}\bar{s}^2
\end{equation}
Accordingly, the K\"{a}hler form $\omega$ can be cast as,
\begin{equation}\label{kahler1}
\omega =\frac{i}{2}\big{(}1+\epsilon f_1(s^1)\big{)}
\mathrm{d}s^1\wedge\mathrm{d}\bar{s}^1+\frac{i}{2}\big{(}1+\epsilon
f_2(s^2)\big{)} \mathrm{d}s^2\wedge\mathrm{d}\bar{s}^2
\end{equation}
In addition, we further assume that $f_1(s^1)$ and $f_2(s^2)$
vanish at $s^1=0$ and $s^2=0$ respectively, so that the complex
curve $\Sigma$ can be formally defined as a divisor of the complex
surface $S$, and consequently we do not alter the definition of
$\Sigma$. Moreover, the functions $f_1,f_2$ are assumed to be
holomorphic functions of their coordinates, in order the solutions
of the equations of motion are sections of holomorphic line
bundles along the loci $s^2=0$ \cite{fzero,fzero1,fzero2}. We
shall not be interested in the particular form of the solutions
the equations of motion, since we are interested only in
perturbations of the Witten index. Furthermore, we assume that the
functions $f_1$ and $f_2$ are decreasing functions of their
arguments. The form of the localized solutions around the complex
matter curve $\Sigma$ will have a more evolved form in reference
to the non-perturbed metric ones. Let us examine the perturbation
around the complex curve $\Sigma$, that is around $s^2=0$. The
equations of motion corresponding to the $s^2=0$ case, are
(ignoring again some irrelevant constants):
\begin{align}\label{eqmotperturbed1}
& \big{(}1+\epsilon
f_1(s^1)\big{)}\partial_2\psi^a
+\bar{s}^2\chi^a =0
\\  \notag &
{\,}{\,}{\,}{\,}{\,}{\,}{\,}{\,}{\,}{\,}{\,}{\,}{\,}{\,}{\,}{\,}{\,}{\,}{\,}{\,}{\,}{\,}{\,}{\,}{\,}{\,}{\,}{\,}{\,}{\,}{\,}{\,}{\,}{\,}{\,}{\,}{\,}{\,}{\,}{\,}{\,}{\,}\bar{\partial}_2\chi^a+s^2\psi^a
=0
\end{align}
which easily comes to be,
\begin{align}\label{eqmotperturbed12}
& \partial_2\psi^a
+\frac{\bar{s}^2}{\big{(}1+\epsilon
f_1(s^1)\big{)}}\chi^a =0
\\  \notag &
{\,}{\,}{\,}{\,}{\,}{\,}{\,}{\,}{\,}{\,}{\,}{\,}{\,}{\,}{\,}{\,}{\,}{\,}{\,}{\,}{\,}{\,}{\,}{\,}{\,}\bar{\partial}_2\chi^a+s^2\psi^a
=0
\end{align}
Expanding in powers of $\epsilon$ and keeping linear terms, we
get:
\begin{align}\label{eqmotperturbed14}
& \partial_2\psi^a +\bar{s}^2\big{(}1-\epsilon
f_1(s^1)\big{)}\chi^a =0
\\  \notag &
{\,}{\,}{\,}{\,}{\,}{\,}{\,}{\,}{\,}{\,}{\,}{\,}{\,}{\,}{\,}{\,}{\,}{\,}{\,}{\,}{\,}{\,}{\,}{\,}{\,}{\,}{\,}{\,}{\,}{\,}{\,}{\,}{\,}{\,}{\,}{\,}{\,}{\,}{\,}{\,}{\,}{\,}\bar{\partial}_2\chi^a+s^2\psi^a
=0
\end{align}
The zero modes of equation (\ref{eqmotperturbed14}), are the zero
modes of the operator:
\begin{equation}\label{matrixodd}
\mathcal{D}_{\epsilon}=\left(%
\begin{array}{cc}
  \partial_2 & \bar{s}^2\big{(}1-\epsilon
f_1(s^1)\big{)} \\
  s^2 & \bar{\partial}_{2} \\
\end{array}%
\right)
\end{equation}
We can write $\mathcal{D}_{\epsilon}=\mathcal{D}_F+\mathcal{K}$, with $\mathcal{D}_F$ as in equation
(\ref{susyqmrn567m}) and $\mathcal{K}$ is the operator:
\begin{equation}\label{codd}
\mathcal{K}=\left(%
\begin{array}{cc}
  0 & \bar{s}^2\epsilon
f_1(s^1) \\
  0 & 0 \\
\end{array}%
\right)
\end{equation}
We know from the mathematical literature, that compact
perturbations of the index of Fredholm operators, leave the index
invariant, that is:
\begin{itemize}
\item[*]For $Q$ be a Fredholm operator and $\mathcal{K}$ be a compact odd
operator, then operator $Q+\mathcal{K}$ is a Fredholm operator and also:
\begin{equation}\label{indexfredtheorodd}
\mathrm{ind}(Q+\mathcal{K})=\mathrm{ind}Q
\end{equation}
\end{itemize}
The operator $\mathcal{K}$ is a compact odd operator. Odd, since it
anti-commutes with $W$ and compact since we assumed that the
functions $f_i$ are rapidly decreasing functions of their
arguments. Thereby we have the result that:
\begin{equation}\label{indexfredtheorodrgtd}
\mathrm{ind}(\mathcal{D}_F+\mathcal{K})=\mathrm{ind}\mathcal{D}_F
\end{equation}
Hence the Witten index of the operator $\mathcal{D}_{\epsilon}$ is equal to
the Witten index of the operator $\mathcal{D}_F$. We can conclude that the
SUSY QM algebra remains invariant under rapidly decaying
perturbations of the metric. This result is valid, even if we
include higher values of $\epsilon$ in the operator $\mathcal{K}$. This can
be easily verified, since the operator $\mathcal{K}$ would be equal to:
\begin{equation}\label{matrixchigher}
\mathcal{K}=\left(%
\begin{array}{cc}
  0 & \bar{s}^2\epsilon
f_1(s^1)+\bar{s}^2\epsilon^2
f_1^1(s^1)+... \\
  0 & 0 \\
\end{array}%
\right)
\end{equation}
This operator is still a compact odd operator and satisfies the
aforementioned theorem.

\section{Geometrical Implications of the $N=2$ SUSY QM Algebras}

An underlying $N=2$, $d=1$ supersymmetric quantum algebra
related to some fermions of the eight dimensional super-Yang Mills theory, implies some additional geometric structures over $S\times R^{3,1}$, and particularly to the associated fibre bundles, to which fermions are sections.

\noindent It would be helpful to discuss in short what is our
motivation to study such extra geometric structures over $S\times
R^{3,1}$, what is our aim and moreover present briefly the main
results of this study. The main motivation to our study comes from
the fact that some of the sections of the spin bundle over
$S\times R^{3,1}$ (the ones localized on the complex dimension one
curve $\Sigma$), are related to an $N=2$, $d=1$ supersymmetric
quantum algebra, and particularly, the fermions are directly
related to the vectors of the graded Hilbert space corresponding
to the SUSY QM algebra. This fact suggests that the fermionic
sections are directly related to some underlying affine vector
bundle that has an intrinsic graded structure. In turn, this in
conjunction with the fact that only the $\Sigma$-localized
fermions are directly correlated to the SUSY QM algebra, strongly
suggests that the total spin bundle over $S\times R^{3,1}$ is
reducible (but not projectable) to some subbundle that consists of
the affine $Z_2$ graded vector bundle. This reducibility is
materialized in terms of the corresponding connections and
covariant differentials. We shall formally address these issues in
the following subsections. For some references on connections and
covariant differentials see for example
\cite{graded1,graded2,graded3,graded4,graded5,graded6,Jost,Nakahara,eguchi}
while for issues regarding graded manifolds see
\cite{graded,graded1,graded2,graded3,graded4,graded5,graded6}.

\subsection{Geometric Structures for the Fermionic Sector}

In order to simplify our notation, we denote $X$ the space $S\times R^{3,1}$ upon which the SUSY Yang-Mills
model we described in this paper, is built on. The fermionic fields are actually sections of the $G_S-$twisted
fibre bundle $P\times S\otimes G_S$, with $S$ the
 Spin group $Spin(8)$ representation and $P$, the double cover of the
principal $SO(8)$ bundle on the tangent manifold $TX$.

As we already mentioned, the $\Sigma$-localized fermions, are elements of graded the vector space of the supersymmetric
quantum algebra. The graded vector space
$\mathcal{H}=\mathcal{H}^+\oplus \mathcal{H}^-$, that some of the
sections of the total fibre bundle $P\times S\otimes G_S$ belong,
strongly suggests a new geometric structure on the manifold $X$. Particularly $X$
becomes a graded manifold $(X,\mathcal{A})$.

\noindent The $N=2$ SUSY QM algebra,
$\mathcal{W},\mathcal{Q}_{F},\mathcal{Q}_{F}^{\dag}$ and particularly the
involution $W$, generates the $Z_2$ graded vector space
$\mathcal{H}=\mathcal{H}^+\oplus \mathcal{H}^-$. The subspace
$\mathcal{H}^+$ contains $W$-even vectors while $\mathcal{H}^-$, $W$-odd vectors.
This grading is an additional algebraic structure on $X$, with
$\mathcal{A}=\mathcal{A}^+\oplus \mathcal{A}^-$ an $Z_2$ graded
algebra. Formally, $\mathcal{A}$ is a total rank $m$ ($m=2$ for
our case) sheaf of
$Z_2$-graded commutative $R$-algebras. In addition, the sheaf $\mathcal{A}$ underlies the vector
space $\mathcal{H}$ and makes the space $\mathcal{H}$
an $Z_2$-graded $\mathcal{A}$-module. This can be verified
from the fact that:
\begin{equation}\label{amodule}
A_+\cdot H_+\subset H_+,{\,}{\,}A_+\cdot H_-\subset
H_-,{\,}{\,}A_-\cdot H_+\subset H_-,{\,}{\,}A_-\cdot H_-\subset
H_+
\end{equation}
The sheaf $\mathcal{A}$ contains the endomorphism $W$ (which is the
involution of the SUSY quantum algebra), $W:\mathcal{H}\rightarrow
\mathcal{H}$, with $W^2=I$, providing the $Z_2$-grading on the vector space
$\mathcal{H}$, {\it i.e.}:
\begin{equation}\label{gred}
W\mathcal{H}^{\pm}=\pm 1
\end{equation}
Thereby, $\mathrm{End}(\mathcal{H})\subseteq \mathcal{A}$. $\mathcal{A}$ is called a structure sheaf of the graded manifold
$(X,\mathcal{A})$, while $X$ is the body of
$(X,\mathcal{A})$.

\noindent Since we shall make use of connections on graded manifolds, it is useful to see how these are constructed. Given an open neighborhood $U$ of $x$ $\in $ $X$, we have
that locally (bearing in mind that in our case $m=2$):
\begin{equation}\label{sheafloc}
\mathcal{A}(U)=C^{\infty}(U)\otimes\wedge R^m
\end{equation}
Therefore, the structure sheaf $\mathcal{A}$ is isomorphic to the sheaf
$C^{\infty}(U)\otimes\wedge R^m$ of some exterior affine vector bundle
$\wedge \mathcal{H_E}^*=U\times \wedge R^m$, with $\mathcal{H_E}$
the affine vector bundle with fiber the vector space $\mathcal{H}$. The structure sheaf
$\mathcal{A}=C^{\infty}(U)\otimes\wedge \mathcal{H}$, is
isomorphic to the sheaf of sections of the exterior vector bundle
$\wedge \mathcal{H_E}^*=R\oplus
(\oplus^{m}_{k=1}\wedge^k)\mathcal{H_E}^*$. As we shall see, the bundle
$\mathcal{H_E}$ is a crucial component of the composite bundle we shall construct.

\noindent The connections of the graded manifold $(X,\mathcal{A})$
compose an affine space modeled on the space of
sections of the vector bundle $TX^*\otimes \wedge
\mathcal{H_E}^*\otimes \mathcal{H_E}$. The sections of
the bundle $TX^*\otimes \wedge \mathcal{H_E}^*\otimes
\mathcal{H_E}$ are operators that belong to the sheaf
$\mathcal{A}$ with $\mathcal{H}$-valued forms as elements in some
specific representation, the dimension of which depends on the
rank of the sheaf. In the  present case, these are $2\times 2$
matrices with $\mathcal{H}$-valued forms as matrix
elements. The graded connection will serve as an auxiliary object, since we are interested in the composition of this auxiliary connection with
another connection we shall introduce soon.

\noindent The Dirac covariant differential, denoted as $\nabla^{\gamma_s}$ corresponding to
the connection $\gamma_s$ of the total bundle, $P\times S\otimes G_S$, results to the equations (\ref{eqmotion1}) and (\ref{eqmotion12}).
But since only the set of equations (\ref{eqmotion12}) are connected to
an unbroken $N=2$ SUSY QM algebra, suggests that the connection $\gamma_s$ is reducible to
a connection $\gamma_C$, which in some way is related to the
graded manifold $(X,\mathcal{A})$. This reducibility
implies that the total covariant differential of the fibre bundle
$P\times S\otimes G_S$, is reducible to one covariant differential that when it acts on
integral sections of the fibre bundle $P\times S\otimes G_S$ (in this paper we mainly focused on the integral sections of the fibre
bundles), we get the equations of (\ref{eqmotion12}).

\noindent The $N=2$ SUSY QM algebra suggests an
underlying geometric structure that is pictured by the following
diagram:
$$
\harrowlength=130pt \varrowlength=60pt \sarrowlength=120pt
\commdiag{X&\mapright^{\gamma_s}&S\times P\otimes
G_S{\,}{\,}{\,}{\,}{\,}\cr
&\arrow(3,-2)\lft{\gamma_E}&\mapup\rt{\gamma_{SE}}\cr &&
{\,}{\,}{\,}TX^*\otimes \wedge \mathcal{H_E}^*\otimes
\mathcal{H_E}\cr}$$ with the arrows showing the
directions of the connections of the corresponding fibre bundles and not the direction of the
projective maps of the bundles. The connections appearing above are morphisms of the following fibre maps:
\begin{align}\label{mapconn}
& \gamma_s: P\times S\otimes G_S\rightarrow J^1(P\times S\otimes
U(1)),
\\ \notag&(\mathrm{Bundle}{\,}\mathrm{map},{\,}{\,}\pi_s:P\times S\otimes G_S\rightarrow X)
\\ \notag
& \gamma_E: TX^*\otimes \wedge \mathcal{H_E}^*\otimes
\mathcal{H_E} \rightarrow J^1(TX^*\otimes \wedge
\mathcal{H_E}^*\otimes \mathcal{H_E})
\\ \notag & (\mathrm{Bundle}{\,}\mathrm{map},{\,}{\,}\pi_E:TX^*\otimes \wedge \mathcal{H_E}^*\otimes \mathcal{H_E}\rightarrow X) \\ \notag
& \gamma_{SE}: (P\times S\otimes G_S)_G\rightarrow J^1((P\times
S\otimes G_S)_G)
\\ \notag &(\mathrm{Bundle}{\,}\mathrm{map},{\,}{\,}\pi_{SE}: P\times S\otimes G_S\rightarrow TX^*\otimes \wedge \mathcal{H_E}^*\otimes \mathcal{H_E})
\end{align}
In the above, $J^1Y_i$ stands for the jet bundle of the bundle $Y_i$.
The underlying geometrical structure is actually a composite fibre bundle
over the manifold $X$, of the form:
\begin{equation}\label{compbund}
P\times S\otimes G_S\xrightarrow{\pi_{SE}} TX^*\otimes \wedge
\mathcal{H_E}^*\otimes \mathcal{H_E} \xrightarrow{\pi_{E}} X
\end{equation}
The composite connection corresponding to the
composite fibre bundle (\ref{compbund}) is defined as follows:
\begin{equation}\label{compconnection}
\gamma_C=\gamma_{SE}\circ \gamma_E
\end{equation}
which obviously is the composition of the connections
$\gamma_{SE}$ and $\gamma_E$. The composite
connections appear in the first order differential
operators and the corresponding covariant differentials, thus altering their final form. It worths remembering the definition of the first order differential and of the covariant differential corresponding to
a connection. Let $Y\rightarrow X$, be an arbitrary fibre bundle, with $s_Y: X\rightarrow Y$ a section, and a connection $\gamma_Y: Y\rightarrow J^1 Y$. The first order differential is defined to be:
\begin{equation}\label{fod}
\mathcal{D}_{\gamma_Y}:J^1 Y\rightarrow TX^*\otimes VY
\end{equation}
In the above relation, $VY$ denotes the vertical subbundle of $Y$, which when $Y$
is a vector bundle, we have $VY=Y\times Y$. Accordingly, the covariant differential
corresponding to $\gamma_Y$, denoted $\nabla^{\gamma_Y}$, is:
\begin{equation}\label{covdiff}
\nabla^{\gamma_Y}=\mathcal{D}_{\gamma_Y}\circ J^1 s_Y:
X\rightarrow TX^*\otimes VY
\end{equation}
Therefore, the total covariant differential of the bundle $P\times
S\otimes G_S\rightarrow X$, is equal to:
\begin{equation}\label{totcovdif}
\nabla^{\gamma_{s}}=\mathcal{D}_{\gamma_{s}}\circ J^1 s:
X\rightarrow TX^*\otimes V(P\times S\otimes G_S)\equiv
TX^*\otimes P\times S\otimes G_S
\end{equation}
When this covariant differential acts on integral sections of $P\times
S\otimes G_S\rightarrow X$,
results to equations
(\ref{eqmotion1}) and (\ref{eqmotion12}). Before proceeding, it worths discussing something
very crucial for the rest of the analysis. We denote by $s_E$ and $s_{SE}$
the sections of the following bundles:
\begin{align}\label{mapconn}
& s_E:X\rightarrow TX^*\otimes \wedge \mathcal{H_E}^*\otimes
\mathcal{H_E}
\\ \notag
& s_{SE}: TX^*\otimes \wedge \mathcal{H_E}^*\otimes
\mathcal{H_E}\rightarrow P\times S\otimes G_S
\end{align}
The subbundle $Y_h$, is the restriction $Y_h=s_E^*(P\times S\otimes G_S)$ of the fibre
bundle $\pi_{SE}: P\times S\otimes G_S\rightarrow TX^*\otimes
\wedge \mathcal{H_E}^*\otimes \mathcal{H_E}$, to the submanifold
$s_E(X)\subset P\times S\otimes G_S$, through the inclusion map
\begin{equation}\label{8inc}
i_h:Y_h \hookrightarrow P\times S\otimes G_S
\end{equation}
The composition of the sections $s_E$ and $s_{SE}$ is $s_C=s_{SE}\circ s_E$. The section $s_C$, is
a section of the fibre bundle $\pi_s:P\times S\otimes
G_S\rightarrow X$, with $s_C(X)\subset P\times S\otimes G_S$.
Accordingly, the covariant differential for the composite
bundle corresponding to the connection
(\ref{compconnection}), which we denote $\nabla^{\gamma_{C}}$, follows easily:
\begin{equation}\label{totcovdifcomp}
\nabla^{\gamma_{C}}=\mathcal{D}_{\gamma_{C}}\circ J^1 s_C:
X\rightarrow TX^*\otimes VY_h\equiv TX^*\otimes Y_h
\end{equation}
When the covariant differential is applied to integral sections of
$Y_h$, which form a subset of the set of  integral sections of $P\times
S\otimes G_S$, we obtain the set of equations of relation
(\ref{eqmotion12}). Therefore, we may conclude that the
covariant differential, $\nabla^{\gamma_{s}}$ of the total bundle
$P\times S\otimes G_S$ is reducible to the covariant differential
$\nabla^{\gamma_{C}}$. This implies that the connection
$\gamma_s$ is reducible to $\gamma_C$ (but in any case not
projectable), in such a way so that the following diagram is commuting:
$$\commdiag{P\times S\otimes G_S&\mapright^{\gamma_s}&J^1Y_h\cr
\mapup\lft{i_h}&&\mapup\rt{J^1i_h}\cr
Y_h&\mapright^{\gamma_C}&J^1P\times S\otimes G_S\cr}$$ Recall that the
inclusion map $i_h$ is the one appearing in relation (\ref{8inc}) and
$J^1i_h$ the corresponding jet prolongation of this inclusion map.

\subsection{Sections of Canonical Bundles and Sections of the Composite Bundle}

Using the geometric constructions we just presented, we can directly connect the sections of the associated fibre bundles corresponding to the underlying graded geometry, to the local geometry of the complex curve. Recall that $\Sigma$ is defined to be the complex curve inside $S$ with $s^2=0$. We mentioned earlier that only the localized fields around the curve $\Sigma$ can be associated to the underlying graded structure, but let us give a more formal description of the localized fields around $\Sigma$ in terms of line bundles of the curve. The normal bundle to the curve $\Sigma$, which we denote $\mathcal{N}_{\Sigma/S}$ is parametrized by the coordinate $s^1$ and satisfies:
\begin{equation}\label{normal}
\mathcal{N}_{\Sigma/S}\otimes \mathcal{N}_{\Sigma/S}\cong \mathcal{K}_{\Sigma/S}^{-1}
\end{equation}
with $\mathcal{K}_{\Sigma/S}^{-1}$ the anti-canonical bundle of the manifold $\Sigma$. This obviously implies that the normal bundle $\mathcal{N}_{\Sigma/S}$ is equivalent to $\mathcal{K}_{\Sigma/S}^{-1/2}$, which means that a section of $\mathcal{N}_{\Sigma/S}$, transforms as a anti-holomorphic $1/2$ form on $\Sigma$, that is $(-1/2,0)$. This fact defines a spin structure on $\Sigma$, since the sections of the inverse normal bundle $\mathcal{N}_{\Sigma/S}^{-1}$ are holomorphic $(1/2,0)$ forms on $\Sigma$. This could be deducted directly from equation (\ref{normal}), by exploiting the fact that the canonical bundle $\mathcal{K}_{\Sigma/S}$ in our case is a line bundle, and hence relation (\ref{normal}), reveals a local spin structure around $\Sigma$. So geometrically the localized fermions around $\Sigma$ are actually the sections of the inverse normal bundle $\mathcal{N}_{\Sigma/S}$ or equivalently the sections of the square root of the canonical line bundle (in our case) $\mathcal{K}_{\Sigma/S}^{1/2}$. Using the terminology and notation of the previous section, we can state that the sections $s_C=s_{SE}\circ s_E$ of the composite fibre bundle $\pi_s:P\times S\otimes
G_S \xleftarrow{s_{SE}} TX^*\otimes \wedge \mathcal{H_E}^*\otimes
\mathcal{H_E}\xleftarrow{s_{E}} X$, are in bijective (one-to-one) correspondence to the sections of the square root of the canonical bundle $\mathcal{K}_{\Sigma/S}^{1/2}$, corresponding to the complex curve $\Sigma$.

\section*{Concluding Remarks}

In this paper we studied an eight dimensional super-Yang Mills theory, compactified on $S\times R^{3,1}$. Working locally in $\Sigma \subset S$ we established the result that some sections of the spin bundle over $S\times R^{3,1}$, and particularly those localized on $\Sigma$, are related to an unbroken $N=2$ SUSY QM algebra. This SUSY QM algebra
provides the fermionic sector with an additional geometric structure over
the $S\times R^{3,1}$ space. Particularly, this geometric
structure consists of a composite bundle over the space $S\times R^{3,1}$, with
a graded manifold over it being a basic ingredient of the
composite bundle. Specifically, we demonstrated that the covariant differential $\nabla^{\gamma_{s}}$ is reducible to $\nabla^{\gamma_{C}}$
corresponding to some subbundle $Y_h$ of the initial total $G_S$-twisted spin
bundle. Moreover, we have shown that the sections of the composite fibre bundle $\pi_s:P\times S\otimes
G_S \xleftarrow{s_{SE}} TX^*\otimes \wedge \mathcal{H_E}^*\otimes
\mathcal{H_E}\xleftarrow{s_{E}} X$, are in bijective correspondence to the sections of the square root of the canonical bundle $\mathcal{K}_{\Sigma/S}^{1/2}$, which are the localized zero modes on the curve $\Sigma$.

\noindent The $N=1$ spacetime supersymmetry of the initial theory on $S\times R^{3,1}$, implies that we can construct a supermanifold on $S\times R^{3,1}$. Interestingly enough, apart from the $N=1$ supermanifold which we can construct
over $S\times R^{3,1}$, that manifold is also an $Z_2$-graded manifold, with
the grading structure provided by the involution $W$, the
Witten parity of the SUSY QM system. It worths examining if there is any direct correlation between the
$N=1$ supermanifold and the graded manifold (which is not a
supermanifold) motivated by the fact that every
graded manifold defines a DeWitt supermanifold.

\noindent Moreover, it would be of some importance to look for
higher order one dimensional extended supersymmetries, possibly
with non-zero supercharge \cite{ivanov1,ivanov2,ivanov3}. There is
an indication of such a structure, since the $N=1$ supersymmetry
implies that the zero modes of the fermions are accompanied by
bosonic zero modes. In turn, a SUSY QM algebra can be built for
the bosonic sector (or the corresponding bosonic fluctuations).
Hence it probable that the two $N=2$ SUSY QM algebras can combine
to give a higher order extended $d=1$ supersymmetry.

\noindent Before closing, we have to comment that the results
presented in this article are mostly relevant to mathematical
problems. Particularly, in view of the fact that the complexity of
a generally difficult problem is reduced by finding symmetries or
regularities of a system in general, we established the result
that there is a one-to-one correspondence between the sections of
the composite fibre bundle related the SUSY QM structure and the
sections of the square root of the F-theory canonical bundle. This
is of particular importance since the graded manifold defines
locally a DeWitt supermanifold, and hence we may extend the line
of research we adopted to the global supersymmetric structure of
the theory. Apart from these mathematical applications, there is a
probable physical application. Particularly, the results we
presented might shed some light on the local supersymmetry
structure of the theory, especially if the gravitational
back-reaction to the local metric is taken into account. But these
tasks are outside the scope of this article.

\end{document}